\documentstyle[onecolumn]{mn}
\begin{document}

\voffset=-1.5cm

\title{Radiation drag effects on
magnetically-dominated outflows around compact objects}

\author[V.S.~Beskin, N.L.~Zakamska and H.~Sol]
       {V.S.~Beskin,$^{1}$ N.L.~Zakamska$^{2,3}$ and H.~Sol$^{4}$ \\
           $^1$ P.N.~Lebedev Physical Institute, Leninsky prosp., 53,
Moscow, 117924, Russia\\
           $^2$ Department of Astrophysical Sciences, Princeton University,
           Princeton, NJ, 08544, USA \\
           $^3$ Moscow Institute of Physics and Technology, Dolgoprudny, 141700,
           Russia \\
           $^4$ LUTH, Observatory Paris-Meudon, Pl. J.Janssen 5,
Meudon, 92195, France}
\date{Accepted   .
     Received   ;
     in original form }


\pubyear{2003}

\maketitle


\begin{abstract}
The effects of radiation drag force on the structure of relativistic
electron-positron and electron-proton outflows are considered within
the one-fluid approximation for quasi-monopole
cold outflow. It is shown that for a Poynting-dominated case
the drag force does not change the particle energy inside the fast
magnetosonic surface. In this region the action of the drag results in a
diminishing of the Poynting flux, not the particle flux. Outside the
fast magnetosonic surface, for intermediate photon density the drag force may
result in additional acceleration of the plasma.
This acceleration is a result of the disturbance of magnetic surfaces under the
action of the drag. At even larger distances particles are not frozen into the
magnetic field and the drag force decelerates them efficiently.

In the case of extreme photon densities, the
disturbance of magnetic surfaces becomes large and the drag force
changes the total energy flux significantly, the particles becoming
nonrelativistic.

We find that for Active Galactic Nuclei
the photon density is too low to disturb the parameters of an ideal MHD
outflow. The drag action may result in additional acceleration of
outgoing plasma only for central engines with very high luminosities.
For cosmological gamma-ray bursts the drag force can strongly affect the process of formation of a Poynting-dominated outflow.
\end{abstract}


\section{Introduction}

Magnetohydrodynamic (MHD) models
are now developed intensively in theories of the magnetospheres of
rotating supermassive black holes
($M \sim 10^{8}\hbox{--}10^{9}M_{\odot}$, $B_0 \sim 10^4$
G), which are believed to reside in central engines of
Active Galactic Nuclei
(AGNs) and quasars (Begelman, Blandford \& Rees 1984; Thorne, Price \&
Macdonald 1986). In particular, it is the MHD model that is the most
promising in the problem of the origin and stability of jets. Indeed,
the MHD approach explains both the energy source (the rotational
energy of the compact object) and the mechanism of the energy and
angular momentum loss (for an overview, see e.g. Blandford 2002). Observational
evidence in favor of MHD models was recently found in the possible presence
of toroidal magnetic fields in jets
(Gabuzda et al 1992; Gabuzda et al 1999). Magnetically-dominated
outflows are also believed to be responsible for the energy transport in
cosmological gamma-ray bursts (M\'esz\'aros \& Rees 1997;
Lee, Wijers \& Brown 1998; van Putten \& Levinson 2003),
when energy is released in the merging of black holes or neutron
stars ($M \sim M_{\odot}$, $B_0 \sim 10^{15}$ G).

It has been suggested that the density of photons
in the vicinity of the central engine is so high that they
may drastically change the characteristics of the ideal MHD outflow.
For example, they may result in extensive $e^+e^-$ pair creation
(Svensson 1984),
acceleration of low-energy pairs by the radiation drag force
(Phinney 1982; Turolla, Nobili \& Calvani 1986; Beloborodov 1999)
and deceleration of high-energy particles
(Melia \& K\"onigl 1989; Sikora et al 1996).
In other words, a self-consistent consideration
should take the drag force into account.

So far the two processes  -- the ideal MHD
acceleration and the action of external photons -- have been considered
independently. The first step to combine them was made
by Li, Begelman \& Chiueh (1992). In particular, it was
shown how the equations can be integrated in a conical
geometry (which is impossible in the general case). On the other hand, the
consideration was performed within the approximation of a fixed poloidal
magnetic field. Under this assumption the fast magnetosonic surface of
a cold flow is shifted to infinity (Michel 1969; Kennel, Fujimura \&
Okamoto 1976; Lery et al 1998). As a result, it was impossible to analyze
the effects of radiation drag on the position of the fast magnetosonic
surface and the properties of the outflow outside this surface.

The main goal of this paper is to determine more carefully
the radiation drag effects on a magnetically-dominated
outflow. To describe analytically the effects of radiation drag,
including simultaneously the disturbance of the magnetic surfaces
we consider here a quasi-monopole outflow. For AGNs,
such geometry in the immediate vicinity of the central engine
was recently confirmed by direct observations (Junior, Biretta \& Livio
1999). In other words, in the zeroth approximation (i.e., without drag) we
use the analytical solution for a magnetically-dominated MHD
outflow (Beskin, Kuznetsova \& Rafikov 1998, hereafter Paper I),
in which the fast magnetosonic surface is located at a finite distance from
the origin.

For simplicity we consider the following model of the
radiation field in the vicinity of the central engine. First, we notice that
for ultra-relativistic particles the energy of a photon
propagating nearly along the particle trajectory remains almost the same
after a collision. This means that the drag force from these photons is
small. Thus, only the isotropic component of the photon field contributes
substantially to the drag force.
Hence, in our geometry with a strong central source of photons and a monopole
outflow of particles, only a small fraction of photons
(the isotropic component of the photon field) interacts
efficiently with the particles,
producing inverse Compton photons with energies
${\cal E}_{\rm IC} \sim \gamma^2{\cal E}_{\rm ph}$.

The isotropic component can be produced, firstly, by the outer part of the
accretion disk and, secondly, by external sources. It can be modeled as
\begin{equation}
U = U_{\rm iso} = U_{\rm A}\left(\frac{r}{R_{\rm L}}\right)^{-n}
+ U_{\rm ext}, \label{u_large}
\end{equation}
where $R _{\rm L} = c/\Omega$ is the radius of the light cylinder,
$U_{\rm A} = U(R_{\rm L}) = L_{\rm tot}/(4\pi R_{\rm L}^{2}c)$,
and $n \approx 3$ (for more details see, e.g., Sikora et al 1996).
Here the first term describes the radiation from the outer parts of
the disk, $ r_{\rm rad} > r$,
while the second one corresponds to the homogeneous external
radiation. For AGNs this can be due to clouds located at
a distance $r_{\rm cloud} \sim 1$pc
from the central engine and reradiating $k L_{\rm tot}$ of the total
luminosity ($k \sim 10\%$). In this case
\begin{equation}
U_{\rm ext} = k\frac{L_{\rm tot}}{4\pi r_{\rm cloud}^2c}. \label{u1}
\end{equation}
However, this model only makes physical sense
at distances less than $r_{\rm cloud}$, and the term
vanishes at larger distances.

Finally, as some arguments exist both in favor of
(Reynolds et al 1996; Hirotani et al 1999) and against
(Sikora \& Madejski 2000) the leading role of $e^+e^-$ plasma
in relativistic jets, in what follows we consider both
electron-positron and electron-proton outflows.

In Section 2 we formulate the basic equations describing a
quasi-monopole outflow of relativistic plasma in two-fluid
and one-fluid approximations. Then in Section 3 we analyze
the main properties of an electron-positron outflow. A
similar analysis for electron-proton plasma is produced
in Section 4. Finally in Section 5 we
consider the effects of radiation drag for real astrophysical objects.

\section{Basic Equations}

\subsection{The Two-Fluid Description}

We consider an axisymmetric, stationary outflow of two-component cold
plasma from the magnetosphere of a rotating body with a split monopole
poloidal magnetic field.
This geometry can be realized in the
presence of an accretion disk separating the ingoing and outgoing magnetic
fluxes (Blandford \& Znajek 1977). In the hydrodynamic
approximation, the structure of the flow is described by
Maxwell's equations and the separate equations of motion for
positively and negatively charged particles:
\begin{eqnarray}
\nabla {\bmath E}=4\pi \rho_{\rm e}, \qquad \nabla \times {\bmath E} =0,
\nonumber \\
\nabla {\bmath B}=0, \qquad \nabla \times {\bmath B} =
\frac{4 \pi}{c}{\bmath j}, \label{1} \\
({\bmath v}^{\pm}\nabla){\bmath p}^{\pm}=\pm e\left( {\bmath E}
+ \frac{{\bmath v}^{\pm}}{c}\times{\bmath B}\right) +
{\bmath F}_{\rm drag}^{\pm}. \nonumber
\end{eqnarray}
Here $\bmath E$ and $\bmath B$ are the electric and magnetic fields,
$\rho_{\rm e} = e(n^+ - n^-)$
and ${\bmath j} = e(n^+{\bmath v}^+ - n^-{\bmath v}^-)$
are the charge and current densities,
and ${\bmath v}^{\pm}$ and ${\bmath p}^{\pm}$  are the velocities
and momenta of the charged particles.
Finally, ${\bmath F}_{\rm drag}^{\pm}$ are the
radiation drag forces which, for an isotropic photon field, have the
form
\begin{equation}
{\bmath F}_{\rm drag}^{\pm} = -\frac{4}{3}\frac{{\bmath v}^{\pm}}
{|v^{\pm}|}
\left(\frac{m_{\rm e}}{m_{\pm}}\right)^{2}
\sigma_{\rm T}U_{\rm iso}(\gamma^{\pm})^2.
\end{equation}
Here $\gamma^{\pm}$ are the Lorentz factors, $m_+$ is the mass of the positively charged particles (positrons or protons) and $m_- = m_{\rm e}$ is the mass of the negatively charged particles. $\sigma_{\rm T} = (8\pi/3)(e^2/m_{\rm e}c^2)^2 $ is the Thompson cross section.
To close system (\ref{1}) the continuity equations $\nabla (n^{\pm}{\bmath v}^{\pm}) = 0$ should be added. It is enough to add the equation for one component, e.g.
\begin{equation}
\nabla (n^{+}{\bmath v}^{+}) = 0, \label{1'}
\end{equation}
because the continuity equation for the second component then follows from (\ref{1'}) and the current continuity equation $\nabla{\bmath j} = 0$ (which, in turn, results from Maxwell's equation $\nabla \times {\bmath B} = (4 \pi/c){\bmath j}$).

In the limit of infinite particle energy,
\begin{equation}
\gamma=\infty, \quad
v_r^{(0)}=c, \quad
v_{\varphi}^{(0)}=0, \quad
v_{\theta}^{(0)}=0,\label{2}
\end{equation}
so that
\begin{equation}
\rho_{\rm e} = \rho_s\frac{R^2}{r^2}\cos\theta, \quad
j_r = \rho_sc\frac{R^2}{r^2}\cos\theta,  \quad
j_{\theta} = j_{\varphi} = 0, \label{3}
\end{equation}
the monopole poloidal magnetic field
\begin{equation}
B_r^{(0)} = B_0\frac{R^2}{r^2}, \quad
B_{\theta}^{(0)}=0,\label{4}
\end{equation}
is an exact solution of Maxwell's equations. In this case,
\begin{eqnarray}
B_{\varphi}^{(0)} & = & E_{\theta}^{(0)}
=-B_0\frac{\Omega R}{c}\frac{R}{r}\sin\theta, \\
E_r^{(0)} & = & E_{\varphi}^{(0)}=0, \label{5}
\end{eqnarray}
which is just the well-known Michel (1973) solution.
Here $B_0$ and $\rho_s$ are the magnetic field and charge density
on the surface $r=R \ll R_{\rm L}$, and the angular velocity is
$\Omega = 2\pi c |\rho_s|/B_0$.
The limit $\gamma \rightarrow \infty$ corresponds to zero
particle mass in the force-free approximation.

It is also convenient to introduce the electric field potential
$\Phi(r,\theta)$,
so that ${\bmath E}=-\nabla \Phi$ and
\begin{equation}
\Phi^{(0)} = -\frac{\Omega R^2 B_0}{c}\cos\theta, \label{6}
\end{equation}
and the flux function $\Psi(r,\theta)$, so that the poloidal magnetic
field
\begin{equation}
{\bmath B}_{\rm p} = \frac{\nabla \Psi\times {\bmath e}_{\varphi}}
{2 \pi r\sin\theta}, \label{7}
\end{equation}
and $\Psi^{(0)} = 2\pi B_0 R^2(1-\cos\theta)$.

Then the dimensionless
corrections $\eta^{\pm}(r,\theta)$, ${\bmath \xi}^{\pm}(r,\theta)$,
$\delta(r,\theta)$,
$\varepsilon f(r, \theta)$ and $\zeta(r,\theta)$
for the case $v\ne c$ can be introduced in the
following form:
\begin{eqnarray}
n^{+} & = & \frac{\Omega B_0}{2\pi ce}\frac{R^2}{r^2}
\left[\lambda-\frac{1}{2}\cos\theta+\eta^{+}(r,\theta)\right], \label{8}\\
n^{-} & = & \frac{\Omega B_0}{2\pi c e}\frac{R^2}{r^2}
\left[\lambda+\frac{1}{2}\cos\theta+\eta^{-}(r,\theta)\right], \label{9}\\
v_r^{\pm} & = & c\left[1-\xi_r^{\pm}(r,\theta)\right],
\qquad v_{\theta}^{\pm} = c\xi_{\theta}^{\pm}(r,\theta),
\qquad v_{\varphi}^{\pm} = c\xi_{\varphi}^{\pm}(r,\theta), \label{12}\\
\Phi(r,\theta) & = & \frac{\Omega R^2 B_0}{c}\left[-\cos\theta
+\delta(r,\theta)\right], \label{10}\\
\Psi(r,\theta) & = & 2\pi B_0 R^2\left[1-\cos\theta+\varepsilon
f(r,\theta)\right], \quad \mbox{and thus} \label{11}\\
B_r & = & B_0\frac{R^2}{r^2}\left( 1+\frac{\varepsilon}{\sin\theta}
\frac{\partial f}{\partial \theta}\right), \label{13}\\
B_{\theta} & = & -\varepsilon \frac{B_0 R^2}{r \sin\theta}
\frac{\partial f}{\partial r}, \label{14}\\
B_{\varphi} & = & B_0\frac{R\Omega}{c}\frac{R}{r}\left[-\sin\theta
-\zeta(r,\theta)\right], \label{15}\\
E_r & = & -\frac{\Omega B_0 R^2}{c}\frac{\partial \delta}{\partial r}, \label{16}\\
E_{\theta} & = & \frac{\Omega R^2 B_0}{c r}
\left( -\sin\theta - \frac{\partial \delta}{\partial \theta}\right). \label{17}
\end{eqnarray}
Here $\lambda \gg 1$ is the multiplication parameter
($\lambda=en_s/|\rho_s|$, where
$n_s$ is the number density of particles on the surface $r=R$).
For $\lambda < 1$ the approach under consideration is not valid. In what
follows we consider for simplicity the case $\lambda =$ const. Such a choice
corresponds to a constant particle-to-magnetic flux ratio $\kappa = $ const.

Switching to dimensionless variables, we use below the
dimensionless radius $x=r/R_{\rm L}=r \Omega/c$ and the dimensionless drag
force:
\begin{equation}
F_{\rm d}^{\pm} =\frac{4}{3}\frac{\sigma_{\rm T}U_{\rm iso}}
{\Omega m_{\rm e}c}(\gamma^{\pm})^2.
\label{fd}
\end{equation}
Now, substituting (\ref{8})--(\ref{17}) into (\ref{1}) we
obtain, to first order in all the correcting functions, the following
system of equations:
\begin{eqnarray}
-\frac{1}{\sin\theta}\frac{\partial}
{\partial \theta}(\zeta \sin\theta)= 2(\eta^+-\eta^-)
-2\left[\left(\lambda-\frac{1}{2}\cos\theta\right)\xi_r^+
-\left(\lambda+\frac{1}{2}\cos\theta\right)\xi_r^-\right],
\label{b1}\\
2(\eta^+-\eta^-)+\frac{\partial}{\partial x}\left(x^2 \frac{\partial
\delta}{\partial x}\right)+ \frac{1}{\sin\theta}\frac{\partial}{\partial
\theta} \left(\sin\theta \frac{\partial\delta}{ \partial \theta}\right)=0,
\label{k1} \\
\frac{\partial\zeta}{\partial x}=\frac{2}{x}
\left[\left(\lambda-\frac{1}{2}\cos\theta\right)\xi_{\theta}^+
-\left(\lambda+\frac{1}{2}\cos\theta\right)\xi_{\theta}^-\right],
\label{dz} \\
-\frac{\varepsilon}{\sin\theta}\frac{\partial^2 f}{\partial
x^2} -\frac{\varepsilon}{x^2}\frac{\partial}{\partial\theta}
\left(\frac{1}{\sin\theta}\frac{\partial f} {\partial \theta}\right)=
\frac{2}{x}\left[\left(\lambda-\frac{1}{2}\cos\theta\right)\xi_{\varphi}^+
-\left(\lambda+\frac{1}{2}\cos\theta\right)\xi_{\varphi}^-\right],\\
\frac{\partial}{\partial x}\left(\xi_{\theta}^+\gamma^+\right)+
\frac{\xi_{\theta}^+\gamma^+}{x}= -\xi_{\theta}^+F_{\rm d}^
+\left(\frac{m_{\rm e}}{m_{\rm p}}\right)^3
+4\lambda\sigma\left(\frac{m_{\rm e}}{m_{\rm p}}\right)\left(
-\frac{1}{x}\frac{\partial\delta}{\partial\theta}+
\frac{\zeta}{x}-\frac{\sin\theta}{x}\xi_r^++ \frac{1}{x^2}
\xi_{\varphi}^+\right),
\label{s1}\\
\frac{\partial}{\partial
x}\left(\xi_{\theta}^-\gamma^-\right)+ \frac{\xi_{\theta}^-\gamma^-}{x}=
-\xi_{\theta}^-F_{\rm d}^--4\lambda\sigma\left(
-\frac{1}{x}\frac{\partial\delta}{\partial\theta}+
\frac{\zeta}{x}-\frac{\sin\theta}{x}\xi_r^-+ \frac{1}{x^2}
\xi_{\varphi}^-\right),
\label{tt} \\
\frac{\partial}{\partial
x}\left(\gamma^+\right) = -F_{\rm d}^+\left(\frac{m_{\rm e}}{m_{\rm
p}}\right)^3 +4\lambda\sigma\left(\frac{m_{\rm e}}{m_{\rm p}}\right)\left(
-\frac{\partial\delta}{\partial x}-
\frac{\sin\theta}{x}\xi_{\theta}^+\right), \label{s2}\\
\frac{\partial}{\partial x}\left(\gamma^-\right) =-F_{\rm d}^-
-4\lambda\sigma\left( -\frac{\partial\delta}{\partial x}-
\frac{\sin\theta}{x}\xi_{\theta}^-\right),
\label{g2}\\
\frac{\partial}{\partial x}\left(\xi_{\varphi}^+\gamma^+\right)+
\frac{\xi_{\varphi}^+\gamma^+}{x}= -\xi_{\varphi}^+F_{\rm d}^
+\left(\frac{m_{\rm e}}{m_{\rm p}}\right)^3
+4\lambda\sigma\left(\frac{m_{\rm e}}{m_{\rm p}}\right)\left(
-\varepsilon\frac{1}{x \sin\theta}\frac{\partial f} {\partial
x}-\frac{1}{x^2}\xi_{\theta}^+\right), \\
\frac{\partial}{\partial x}\left(\xi_{\varphi}^-\gamma^-\right)+
\frac{\xi_{\varphi}^-\gamma^-}{x}=
-\xi_{\varphi}^-F_{\rm d}^--4\lambda\sigma\left( -\varepsilon\frac{1}{x
\sin\theta}\frac{\partial f} {\partial
x}-\frac{1}{x^2}\xi_{\theta}^-\right).
\label{b2}
\end{eqnarray}
Here
\begin{equation}
\sigma=\frac{\Omega e B_0 R^2}{4\lambda m_{\rm e} c^3} \gg 1
\label{sigma}
\end{equation}
is the Michel (1969) magnetization parameter describing
the particle-to-electromagnetic energy flux ratio
$W_{\rm part}/W_{\rm em} = (m_{\rm p}/m_{\rm e})\gamma/\sigma$.
Hence, for a Poynting-dominated flow we have $\gamma \ll \sigma$.
As we see, the disturbances of the particle density $\eta^+$ and
$\eta^-$ enter equations (\ref{b1})--(\ref{b2}) only in the combination
$\eta^+-\eta^-$. Therefore, the system (\ref{b1})--(\ref{b2}) is closed,
but the equation of mass continuity (\ref{1'}) is necessary to determine
$\eta^+$ and $\eta^-$ separately.
The system (\ref{b1})--(\ref{b2}) differs from the one considered by
Beskin \& Rafikov (2000, hereafter Paper II)
only by the additional drag terms in the r.h.s. of
(\ref{s1})--(\ref{b2}).

\subsection{The One-Fluid Limit}

Equations (\ref{b1})--(\ref{b2}) describe the flow in the
two-fluid approximation. We now reduce the
complete system of equations (\ref{b1})--(\ref{b2})
to consider the one-fluid approximation.
In the case of an electron-proton outflow there is a small parameter
$m_{\rm e}/m_{\rm p} \sim 10^{-3}$ which allows us to neglect the electron
mass and thus to proceed in a standard way to the one-fluid approximation.
But as demonstrated in Paper II, this can be done in the
electron-positron case for a magnetically-dominated ($\sigma \gg 1$) dense
($\lambda \gg 1$) plasma as well. Because the nonhydrodynamic components
of the velocity are small in this case (cf. Melatos \& Melrose 1996),
\begin{equation}
\frac{\Delta\xi_r^{\pm}}{\xi_r} \sim \lambda^{-1}\sigma^{-2/3} \ , \qquad
\frac{\Delta\xi_{\theta}^{\pm}}{\xi_{\theta}} \sim  \lambda^{-1} \ , \qquad
\frac{\Delta\xi_{\varphi}^{\pm}}{\xi_{\varphi}} \sim
\lambda^{-1}\sigma^{-2/3} \ ,
\end{equation}
we can set $\xi_i^+ =\xi_i^- =\xi_i$ ($i = r, \theta, \varphi$), where $\xi_i$
is the hydrodynamic velocity. In this limit we also have

\begin{equation}
\frac{\delta-\varepsilon f}{\varepsilon f} \sim \lambda^{-2}\sigma^{-2/3}.
\end{equation}
As a result, for a magnetically-dominated outflow with
$\sigma \gg 1$ and $\lambda \gg 1$ in the one-fluid approximation,
\begin{equation}
\delta = \varepsilon f.
\label{delta}
\end{equation}

Finally, equations (\ref{s1}) and (\ref{tt}) together with (\ref{delta})
in the limit $\lambda\sigma \gg 1$ under consideration
give another useful one-fluid relation:
\begin{equation}
-\varepsilon\frac{1}{x}\frac{\partial f}{\partial\theta}+
\frac{\zeta}{x}-\frac{\sin\theta}{x}\xi_r+ \frac{1}{x^2}\xi_{\varphi} = 0,
\label{66}
\end{equation}
this equation being the same as in the drag-free case.
On the other hand, as demonstrated in Paper II,
\begin{equation}
\frac{\xi_{\theta}}{\xi_{\varphi}} \approx \sigma^{-1/3}.
\end{equation}
Hence, in the one-fluid approximation one can set $\xi_{\theta}= 0$
so that
\begin{equation}
\gamma^2=\frac{1}{2\xi_r-\xi_{\varphi}^2}.
\label{33}
\end{equation}

It is necessary to stress that in some cases the monopole geometry
allows one to consider separately the set of equations describing
the particle energy and the set of equations resulting in the
Grad-Shafranov (GS) equation, which determines the disturbance of
magnetic surfaces. Thus we can consider particle energy without
formulating here the general
form of the GS equation. Some asymptotics of the GS equation are
discussed below.

\section{The Electron-Positron Outflow}

\subsection{Integrals of motion}

In this section we consider the properties of the electron-positron
outflow when $m_{\rm p} = m_{\rm e} = m$. In our simple geometry with a
(split) monopole poloidal magnetic field in the zeroth approximation the
particle motion can be considered as radial. It allows us to integrate
some equations along the $r$ coordinate axis. This feature was first
demonstrated by Li et al (1992).

Indeed, combining (\ref{s2}) and (\ref{g2}) with (\ref{dz}) one can obtain
in the one-fluid approximation
\begin{equation}
\zeta = \frac{l(\theta)}{\sin\theta}+\frac{2\varepsilon}{\tan\theta}f
-\frac{1}{\sigma\sin\theta}(\gamma-\gamma_{\rm in})
-\frac{1}{\sigma\sin\theta}l_{\rm A}\int_{x_0}^xu(x^{\prime})
\gamma^2(x^{\prime}) {\rm d}x^{\prime}.
\label{en}
\end{equation}
Here $\gamma_{\rm in}$ is the Lorentz factor near the
origin, $x_{0} = \Omega R/c$, and the integration constant
$l(\theta)$ describes the disturbance of the electric
current $I(R,\theta)=I_A\left[\sin^2\theta+l(\theta)\right]$
on the surface $r = R$.
The integration constant must be determined from the critical conditions on singular surfaces.
The compactness parameter $l_{\rm A}$ is defined by the relation
\begin{equation}
l_{\rm A} = \frac{4}{3}\frac{\sigma_{\rm T}U_{\rm A}}{m_{\rm e}c\Omega},
\label{A}
\end{equation}
where $U_{\rm A} = U(R_{\rm L})$ is again the first term in (\ref{u_large})
on the light cylinder $r= R_{\rm L}$. Finally, we also denote
\begin{equation}
l_{\rm ext} = \frac{4}{3} \frac{\sigma_{\rm T}U_{\rm ext}}{mc\Omega},
\end{equation}
so that (\ref{u_large}) for $x > 1$ can be rewritten as
\begin{equation}
u(x) = x^{-n} + l_{\rm ext}/l_{\rm A}.
\label{m_u}
\end{equation}
For $x < 1$ it is natural to set $u(x) = 1$.

Expression (\ref{en}) is just another way of writing the diminishing of
the Bernoulli integral $E_{\rm B}$ along a magnetic field
line as a result of the drag force. Within the Grad-Shafranov
approach the full energy loss is determined by $W=\int E_{\rm B}{\rm d}\Psi$.
Thus relation (\ref{en}) can be rewritten as
\begin{equation}
E_{\rm B}(r) = E_{\rm B}(R)
- \frac{\lambda\Omega}{2\pi e}\:\int_R^r F_{\rm drag}{\rm d} r,
\label{E}
\end{equation}
where the energy flux per unit magnetic flux $E_{\rm B}$ has the
standard form $E_{\rm B} = \Omega I/2\pi c + \gamma mc^2\kappa$. Here $I$ is
the total electric current inside the magnetic tube and $\kappa$ is the
particle-to-magnetic flux ratio (see e.g. Beskin 1997 for details).

In what follows we consider the case $\gamma_{\rm in} \sim 1$, i.e.
\begin{equation}
\gamma_{\rm in}^3 \ll \sigma,
\label{sgm}
\end{equation}
when additional acceleration of particles
inside the fast magnetosonic surface takes place (see e.g. Paper I). In
the case $\gamma_{\rm in}^3 \gg \sigma$ corresponding to ordinary pulsars,
the particle energy remains constant ($\gamma = \gamma_{\rm in}$) on any
way up to the fast magnetosonic surface (Bogovalov 1997).

The other two integrals of motion, namely
the conservation of angular momentum separately for electrons and
positrons,
can be obtained from equations (\ref{s2})--(\ref{b2}):
\begin{eqnarray}
\gamma^+(1-x\sin\theta\xi_{\varphi}^+)
= \gamma_{\rm in}^+ - l_{\rm A}\int_{x_0}^xu(x^{\prime})
(1-x^{\prime}\sin\theta\xi_{\varphi}^+)(\gamma^+)^2{\rm d}x^{\prime}
-4\lambda\sigma(\delta - \varepsilon f),
\label{mom1}  \\
\gamma^-(1-x\sin\theta\xi_{\varphi}^-)
= \gamma_{\rm in}^- - l_{\rm A}\int_{x_0}^xu(x^{\prime})
(1-x^{\prime}\sin\theta\xi_{\varphi}^-)(\gamma^-)^2{\rm d}x^{\prime}
+4\lambda\sigma(\delta - \varepsilon f).
\label{mom2}
\end{eqnarray}
In the one-fluid approximation this gives
\begin{equation}
\gamma(1-x\sin\theta\xi_{\varphi})=\gamma_{\rm in}
-l_{\rm A}\int_{x_0}^xu(x^{\prime})
(1-x^{\prime}\sin\theta\xi_{\varphi})\gamma^2(x^{\prime}){\rm d}x^{\prime}.
\label{ee}
\end{equation}
At large distances, where $\gamma \gg \gamma_{\rm in}$, we have
\begin{equation}
\xi_{\varphi} \approx \frac{1}{x \sin \theta}.
\label{xiph}
\end{equation}
It can be seen from (\ref{ee}) that the presence of the drag force
(the term proportional to $l_{\rm A}$) makes this approximation more accurate.

\subsection{Fast Magnetosonic Surface}

Substituting $\zeta$ from (\ref{en}),
$\xi_r$ from (\ref{33}), and $\xi_{\varphi}$ from (\ref{ee}) into
(\ref{66}) we get the following equation to determine the position
of the fast magnetosonic surface $r = r_{\rm F}$:
\begin{eqnarray}
-\varepsilon\frac{\partial f}{\partial\theta}
+\frac{2\varepsilon}{\tan\theta}f +
\frac{l(\theta)}{\sin\theta}-\frac{1}{\sigma\sin\theta}
\left(\gamma-\gamma_{\rm in}+J\right) -
\sin\theta\left[\frac{1}{2\gamma^2}+\frac{\left(\gamma-\gamma_{\rm in}
+ J_1\right)^2}{2x^2\gamma^2\sin^2\theta}\right] +
\frac{\gamma-\gamma_{\rm in}+J_1}{x^2 \gamma\sin\theta}=0.
\end{eqnarray}
Here
\begin{eqnarray}
  J & = & l_{\rm A}\int_{x_0}^x u(x^{\prime})
\gamma^{2}(x^{\prime}){\rm d}x^{\prime},
\label{JJJ} \\
J_1 & = & l_{\rm A}\int_{x_0}^x u(x^{\prime})\gamma^{2}(x^{\prime})
(1-x^{\prime}\sin\theta\xi_{\varphi})
{\rm d}x^{\prime}.
\end{eqnarray}
But, in accordance with (\ref{ee}) and thereafter,
the terms $J_1$ and $\gamma_{\rm in}$ can be neglected
far from the origin of the flow $r \gg R_{\rm L}$.
Also, since $l(\theta) \sim \sigma^{-4/3}$ (see Paper I),
the term with $l(\theta)$ can be omitted as well. Hence, one can write down
the following algebraic equation for the Lorentz factor $\gamma$:
\begin{equation}
\gamma^3-\sigma\left(P+\frac{1}{2x^2}\right)\gamma^2
+\frac{1}{2}\sigma\sin^2\theta = 0,
\label{m_gamma}
\end{equation}
where
\begin{equation}
P = - \frac{J}{\sigma} + 2\varepsilon f\cos\theta
-\varepsilon\sin\theta\frac{\partial f}{\partial\theta}.
\label{m_P}
\end{equation}
This differs from the drag-free case by the additional term $J/\sigma$.

Equation (\ref{m_gamma}) allows us to determine the position of the fast
magnetosonic surface $r = r_{\rm F}$ and the energy of particles on this
surface
$\gamma_{\rm F} = \gamma(r_{\rm F})$. Indeed, as the fast magnetosonic surface
is the $X$--point, we find the exact solution for coinciding roots
\begin{equation}
\gamma_{\rm F} = \sigma^{1/3}\sin^{2/3}\theta,
\label{add1}
\end{equation}
which does not depend at all on $P$, or thus on the drag force.
On the other hand,
both the numerator and the denominator of the derivative
${\rm d}\gamma/{\rm d}x$ must be equal to zero on the surface $r=r_{\rm F}$:
\begin{equation}
x\frac{{\rm d}\gamma}{{\rm d}x}=
\frac{\gamma\sigma\left(x{\rm d}P/{\rm d}x-x^{-2}\right)}
{3\gamma-\sigma\left(2P+x^{-2}\right)}.
\end{equation}
It gives
\begin{equation}
(P+x^{-2})_{\rm F} \approx  \sigma^{-2/3}, \qquad
|P|_{\rm F} \approx |x^{-2}|_{\rm F}.
\end{equation}
Now using (\ref{m_P}), one can see that the conditions for weak
and strong drag are, respectively, $l_{\rm A} \ll l_{\rm cr}$
and $l_{\rm A} \gg l_{\rm cr}$, where
\begin{equation}
l_{\rm cr} = \sigma^{1/3}.
\end{equation}
In other words, for $l_{\rm A} \ll l_{\rm cr}$ the flow remains the same
as in the drag-free case, and
\begin{eqnarray}
x_{\rm F} & \approx & \sigma^{1/3}\sin^{-1/3}\theta,
\label{add2} \\
(\varepsilon f)_{\rm F} & \approx & \sigma^{-2/3}.
\label{add3}
\end{eqnarray}
On the other hand, for a high enough photon
density $l_{\rm A} \gg l_{\rm cr}$ one can obtain, for $n = 3$,
\begin{eqnarray}
x_{\rm F} & \approx & \left(\frac{\sigma}{l_{\rm A}}\right)^{1/2}
= \sigma^{1/3}\left(\frac{l_{\rm cr}}{l_{\rm A}}\right)^{1/2},
\label{add2'} \\
(\varepsilon f)_{\rm F} & \approx & \frac{l_{\rm A}}{\sigma}.
\label{var}
\end{eqnarray}
Thus, as we can see, the energy $\gamma_{\rm F}mc^2$ do not depend on the
drag. The disturbance of the magnetic surfaces increases with increasing
$l_{\rm A}$, but remains small for $l_{\rm A} \ll \sigma$.

For a very dense photon field, according to
(\ref{en}) and (\ref{var}), the disturbance of the magnetic surfaces
becomes of order unity for $l_{\rm A} \sim l_{\rm max}$ where
\begin{equation}
l_{\rm max} = \sigma,
\label{lmax}
\end{equation}
i.e.,
when the disturbance of the Bernoulli integral
$\Delta E_{\rm B} = E_{\rm B}(r) - E_{\rm B}(R_{\rm L})$ (\ref{E}) is on
the order of the total energy flux $E_{\rm B}$, or in other words when
the drag force substantially
diminishes the total energy flux of the flow. But this means that
the disturbance of the magnetic surfaces becomes large
when the energies of particles in the vicinity
of the fast magnetosonic surface become nonrelativistic.
This feature is well known for a drag-free flow
both within numerical (Sakurai 1985; Bogovalov 1997) and analytical
(Bogovalov 1992; Tomimatsu 1994; Paper I) considerations.
Of course, our analysis within the small disturbance approach
can only demonstrate the tendency. For this reason, within our approach
$\sigma = $ const. In reality, since $\sigma = E_{\rm B}/(mc^2 \kappa)$,
a decrease of the energy flux results in a decrease of $\sigma$ as well.

Finally, it is necessary to stress that the results depend significantly
on the assumed model of the isotropic photon density.
In particular, for the general power-law
dependence of the photon density for $n < 3$ in (\ref{m_u}), one can obtain
for the position of the fast magnetosonic point and
the disturbance of the magnetic surfaces
\begin{eqnarray}
x_{\rm F} & \approx &
\sigma^{1/3}\left(\frac{l_{\rm cr}}{l_{\rm A}}\right)^{1/(5-n)},\\
(\varepsilon f)_{\rm F} & \approx &
\left(\frac{l_{\rm A}}{\sigma}\right)^{2/(5-n)}
\label{gen}
\end{eqnarray}
instead of (\ref{add2'}) and (\ref{var}).
This takes place for $l_{\rm A} > l_{\rm cr}$, where $l_{\rm cr}$ is the
photon density capable of disturbing the magnetically-dominated outflow:
\begin{equation}
l_{\rm cr} = \sigma^{(n-2)/3}.
\label{acr}
\end{equation}
A large disturbance $(\varepsilon f)_{\rm F}\sim 1$ can only be realized
for a very high photon density $l_{\rm A}>l_{\rm max}$, where again
\begin{equation}
l_{\rm max} = \sigma.
\label{lmax1}
\end{equation}
For $n \ge 3$ (when the value J defined by (\ref{JJJ}) is determined by the
lower limit of integration and does not depend on $n$)
we have $(\varepsilon f)_{\rm F} \approx l_{\rm A}/\sigma$,
$x_{\rm F} \approx (\sigma/l_{\rm A})^{1/2}$,
$l_{\rm cr} = \sigma^{1/3}$, and $l_{\rm max} = \sigma$, i.e. same as for $n=3$.
According to
(\ref{en}), $(\varepsilon f)_{\rm F} \sim 1$ occurs when
$\Delta E_{\rm B}/E_{\rm B} \sim 1$
for all values of $n$.

\subsection{The Structure of the Flow at Small and Large Distances}

\subsubsection{Inner region}

We first consider the structure of the flow well within
the fast magnetosonic surface $r \ll r_{\rm F}$.
As one can see from (\ref{m_gamma}), for $r \ll r_{\rm F}$
\begin{equation}
\gamma \approx \frac{\sin\theta}{\left(2P +1/x^2\right)^{1/2}}.
\end{equation}
In particular, for $l_{\rm A} \ll l_{\rm cr}$ one can neglect
the term $P$ in the denominator. On the other hand, for
$l_{\rm A} \gg l_{\rm cr}$ in the immediate vicinity of the fast
magnetosonic surface the negative term $-J/\sigma$ in $P$
should to be included into consideration. As a result, the particle energy increases abruptly up to the value $\sim \sigma^{1/3}$.
Nevertheless, for $r \ll r_{\rm F}$ the value $P$ can be omitted
for arbitrary compactness parameter $l_{\rm A}$.
It also means that for $r \ll r_{\rm F}$
it is possible to neglect the first two terms in (\ref{66}).
Thus, in the internal region $r \ll r_{\rm F}$ we have
\begin{equation}
\xi_r = \frac{\xi_{\varphi}}{x\sin\theta}.
\label{xir}
\end{equation}
Now using relations (\ref{xiph}) and (\ref{xir}), one can obtain
\begin{eqnarray}
\gamma^2 & = & \gamma_{\rm in}^2+x^2\sin^2\theta
\approx x^2\sin^2\theta,
\label{gamma}\\
\xi_{\varphi} & = & \frac{\sqrt{\gamma_{\rm in}^2+x^2\sin^2\theta}
-\gamma_{\rm in}} {x\sin\theta\sqrt{\gamma_{\rm in}^2+x^2\sin^2\theta}}
\approx
\frac{1}{x\sin\theta},
\label{v1} \\
\xi_r & = & \frac{\sqrt{\gamma_{\rm in}^2+x^2\sin^2\theta}-\gamma_{\rm in}}
{x^2\sin^2\theta\sqrt{\gamma_{\rm in}^2+x^2\sin^2\theta}}
\approx
\frac{1}{x^2\sin^2\theta},
\label{v2}
\end{eqnarray}
in full agreement with the ideal MHD approximation. Hence,
one can conclude that {\it in the internal region} $r < r_{\rm F}$
{\it the radiation drag does
not affect the particle motion}. Here the universal value of the
Lorentz factor (\ref{gamma}) corresponds to the drift velocity,
and it depends on no external disturbances.
Indeed, using the frozen-in condition
${\bmath E} + {\bmath v} \times {\bmath B}/c = 0$,
we obtain for the drift velocity
\begin{equation}
U_{\rm dr}^2 = c^2\frac{{\bmath E}^2}{{\bmath B}^2} =
c^2\left(\frac{B_{\varphi}^2}{E^2}+\frac{B_r^2}{E^2}\right)^{-1}.
\end{equation}
In our case, however, according to (\ref{10})--(\ref{17}),
we have
\begin{equation}
B_{\varphi}^2 \approx E^2,  \qquad
B_r^2 \approx E^2/(x^{2}\sin^{2}\theta).
\end{equation}
These relations immediately lead
to the previously-stated asymptotic behavior (\ref{gamma}). In
particular, this means that disturbance of the magnetic surfaces plays no
role in the determination of the particle energy. For this reason, it is
possible not to consider the radiation drag corrections to the field
structure for $r \ll r_{\rm F}$.

\subsubsection{Outer region}

In the other limit, well outside the fast magnetosonic surface ($r \gg
r_{\rm F}$) equations (\ref{en}), (\ref{66}), (\ref{b1})--(\ref{k1}),
and (\ref{33})
can be rewritten in the form
\begin{eqnarray}
\zeta = \varepsilon\frac{2}{\tan\theta}f -\frac{1}{\sigma\sin\theta}\gamma
-\frac{1}{\sigma\sin\theta}l_{\rm A}\int_{x_0}^xu(x^{\prime})
\gamma^2(x^{\prime}) {\rm d}x^{\prime},
\label{rf1} \\
\zeta = \varepsilon \frac{\partial f}{\partial\theta}+\sin\theta\xi_r,
\label{rf2} \\
\varepsilon\frac{\partial}{\partial x}\left(x^2\frac{\partial f} {\partial x}
\right) - \frac{1}{\sin^3\theta}
\frac{\partial}{\partial\theta}\left(\sin^4\theta\xi_r\right) = 0,
\label{rf3} \\
\gamma^2=\frac{1}{2\xi_r}.
\label{rf4}
\end{eqnarray}
Here we neglect the small values $l(\theta)$, $\gamma_{\rm in}/\gamma$ and
$\xi_{\varphi}$, and set $\delta = \varepsilon f$.
As we can see, the only correction compared to the
drag-free case is the last additional term in (\ref{rf1}).
Without the drag force the system (\ref{rf1})--(\ref{rf4}) results in a
very slow increase of the particle energy ($\gamma \propto \ln^{1/3}r$)
and actually in the absence of collimation,
$\varepsilon f \sim \sigma^{-2/3}\ln^{1/3}r$ (Tomimatsu 1994; Paper I).
Unfortunately, because of the nonlinearity of the system
(\ref{rf1})--(\ref{rf4}), in the general case it is impossible
to reduce it to the GS equation for the magnetic disturbance $\varepsilon
f$ only. For this reason, in what follows we present only an asymptotic
representation of the GS equation.

{\bf a) Low photon density $l_{\rm A} \ll l_{\rm cr}$.}

In this case the action of drag results in a small correction
to the particle energy. Only at very large distances,
where the first term in (\ref{m_u}) can be neglected,
does the drag significantly change the energy of particles.
Indeed, for the decreasing component of the isotropic photon
density ($u(x) \sim x^{-n}$ in (\ref{m_u})) and for an almost
constant particle energy the drag term in (\ref{rf1}) does not increase
with the distance $r$. Here the increase of the drag term is
due to the homogeneous part of the photon density
($l_{\rm ext}/l_{\rm A}$ in (\ref{m_u})).
Clearly, this can be realized at large enough distances
($r > R_{\rm L}(l_{\rm A}/l_{\rm ext})\sigma^{(3-n)/3}$),
where the contribution from external photons is the leading one.
Thus, we see that the action of the drag force
can be significant for a high enough density
of external isotropic photons in the vicinity of the compact
object.

Now we consider this asymptotic region in more detail. Neglecting the term
containing $\gamma$ in (\ref{rf1}) and the first term in (\ref{m_u}),
one can obtain for the generalized GS equation
\begin{equation}
\varepsilon\frac{\partial}{\partial x}\left(x^2\frac{\partial f}
{\partial x}\right)
-\frac{1}{\sin^3\theta}\frac{\partial}{\partial \theta}
\left[
2\varepsilon\sin^2\theta\cos\theta f
-\varepsilon\sin^3\theta\frac{\partial f}{\partial \theta}
-\frac{\sin^2\theta}{\sigma} l_{\rm ext}
\int_{x_0}^x \gamma^2(x^{\rm \prime}){\rm d}x^{\prime}
\right]
= 0.
\label{GS}
\end{equation}
One can seek the solution of this equation in the form
\begin{equation}
\varepsilon f(x,\theta) \propto   x^{\alpha_\delta} \ ,  \qquad
\gamma(x,\theta) \propto x^{\alpha_\gamma}.
\label{ansz}
\end{equation}
Substituting these expressions into
(\ref{GS}) we obtain
\begin{eqnarray}
\varepsilon f(x,\theta) & = &
k_1(\theta)l_{\rm ext}^{1/2}\sigma^{-1/2}x^{1/2},
\label{as1} \\
\gamma(x,\theta) & = & k_2(\theta)l_{\rm ext}^{-1/4}\sigma^{1/4}x^{-1/4},
\label{as2}
\end{eqnarray}
where $k_1(\theta) \sim k_2(\theta) \sim 1$ describe the $\theta$ dependence.
This takes place for $r > r_{\rm h}$, where
\begin{equation}
r_{\rm h} = R_{\rm L}l_{\rm ext}^{-1}\sigma^{-1/3}.
\end{equation}
For $r_{\rm F} < r < r_{\rm h}$ the action of drag is negligible.

As we see, for $r > r_{\rm h}$ the drag force results in a
decrease of the particle energy and an additional collimation
of the magnetic surfaces outside the fast magnetosonic surface.
These asymptotic solutions, however, do not seem to be realized in reality.
Indeed, as one can check, $r_{\rm h} \sim (10^6 - 10^8)R_{\rm L}
> r_{\rm cloud}$. But for $r>r_{\rm cloud}$ the model that we adopted
for the homogeneous component of the photon density is no longer
valid and thus the asymptotic solutions  (\ref{as1})--(\ref{as2}) do not apply.
This conclusion is made under the assumption that the
homogeneous component is produced by the reradiating clouds surrounding the
central engine. One could imagine producing a homogeneous component of the
photon density in the host galaxy. However, in this case the photon density
$U_{\rm ext}$ is too low and thus the collimation distance $r_{\rm h}$ is
too large for our approach to be valid in the region $r\sim r_{\rm h}$.

The characteristic radial dependence of the particle energy in the
presence of the drag is demonstrated in Fig. $1a$. One can conclude that for a low photon density {\it the action of the drag force is very weak unless photon field is present out to distances $r\sim r_{\rm h}$.} In the latter case the drag force efficiently reduces the particle energy beyond $r\sim r_{\rm h}$.

\begin{figure}
\vspace{5.5cm}
\includegraphics{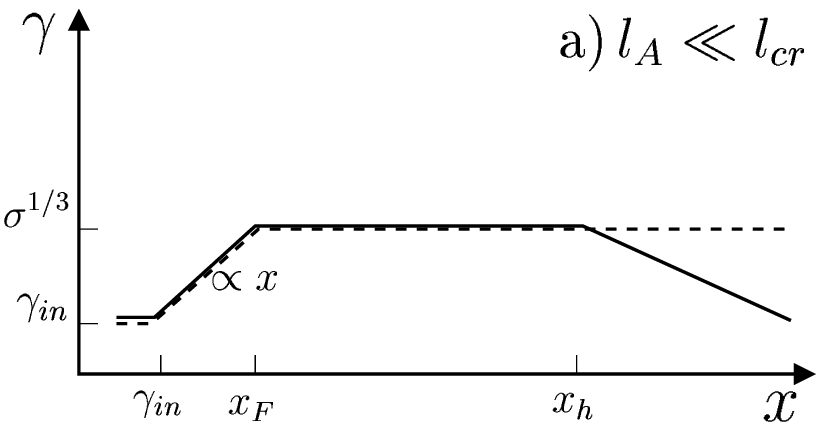}
\includegraphics{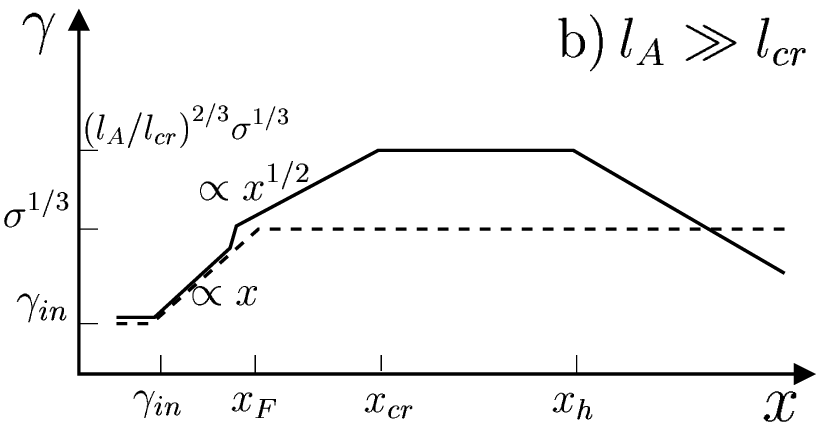}
\caption
{The radial dependence of the Lorentz factor $\gamma(r)$
for low ($a$) and high ($b$) photon densities for $n =3$.
The dashed lines correspond to the drag-free case.}
\end{figure}

It is necessary to stress here an important
property of the fast magnetosonic surface for $l_{\rm A} < l_{\rm cr}$.
Introducing a small disturbance of the
particle energy resulting from the drag
\begin{equation}
\gamma = \gamma_0 + \Delta\gamma,
\label{dg}
\end{equation}
where $\gamma_0$ is the Lorentz factor of particles without
the drag force, one can obtain from (\ref{rf1})--(\ref{rf4})
\begin{equation}
- \sin^2\theta\frac{\Delta\gamma}{\gamma_0^3}
+ \varepsilon\frac{\partial\Delta f}{\partial\theta}
- 2\varepsilon\cos\theta\Delta f
+ \frac{1}{\sigma}\Delta\gamma  = \frac{J}{\sigma}.
\end{equation}
As equation (\ref{rf3}) results in
$\varepsilon \Delta f \sim \Delta\gamma/\gamma_0^3$, we see that
outside the fast magnetosonic surface (where $\gamma_0 > \sigma^{1/3}$)
\begin{equation}
mc^2\Delta\gamma \approx -\int{F_{\rm drag}\, {\rm d}r}.
\end{equation}
On the other hand, within the fast magnetosonic surface we have
\begin{equation}
mc^2|\Delta\gamma| \ll \int{F_{\rm drag}\, {\rm d}r},
\end{equation}
in full agreement with (\ref{gamma}). This means that
inside the fast magnetosonic surface the decrease of the total energy flux
$E_{\rm B}$ results from the decrease of the electromagnetic energy flux rather
than the particle flux. Outside the fast surface the photons act
upon the particle directly. In other words, for $l_{\rm A} < l_{\rm cr}$
the fast magnetosonic surface separates the regions in space where
particles are strongly or weakly frozen into the electromagnetic
field.

{\bf b) High photon density $l_{\rm A} > l_{\rm cr}$.}

For a high photon density $l_{\rm A} > l_{\rm cr}$
the drag force significantly changes the energy of the outgoing particles.
As is shown in Fig. $1b$, the increase of the particle energy continues
outside the fast magnetosonic surface up to the maximum value
\begin{equation}
\gamma_{\rm max} \approx
\left(\frac{l_{\rm A}}{l_{\rm cr}}\right)^{4(n-2)/(5n-n^2)}\sigma^{1/3}.
\label{gmax}
\end{equation}
For $n < 2$ there is no additional acceleration. Only at larger
distances ($r > r_{\rm cr}$) does the radial dependence of the energy
of the outgoing plasma become similar to the previous case.

Indeed, outside the fast
magnetosonic surface one can neglect the term
$\sin\theta\xi_r$ in (\ref{rf2}).
As a result, the system (\ref{rf1})--(\ref{rf4}) can be rewritten
in the form
\begin{eqnarray}
\varepsilon \frac{\partial f}{\partial\theta}
-\varepsilon \frac{2}{\tan\theta}f  =
-\frac{1}{\sigma\sin\theta}\gamma
-\frac{1}{\sigma\sin\theta}l_{\rm A}\int_{x_0}^xu(x^{\prime})
\gamma^2(x^{\prime})
{\rm d}x^{\prime},
\label{qs1} \\
\varepsilon\frac{\partial}{\partial x}\left(x^2\frac{\partial f}
{\partial x}\right) -
\frac{1}{2\sin^3\theta}
\frac{\partial}{\partial\theta}\left(\frac{\sin^4\theta}{\gamma^2}\right) = 0,
\label{qs2}
\end{eqnarray}
which results in
\begin{equation}
\frac{\partial}{\partial x}
\left(x^2\frac{\partial \gamma}{\partial x}\right)
+l_{\rm A}
\frac{\partial}{\partial x}
\left(x^{2-n}\gamma^2\right)
=\frac{\sigma}{2\sin^3\theta}
\left[
2\cos\theta
\frac{\partial}{\partial \theta}
\left(\frac{\sin^4\theta}{\gamma^2}\right)
-\sin\theta
\frac{\partial^2}{\partial \theta^2}
\left(\frac{\sin^4\theta}{\gamma^2}\right)\right] \ .
\label{new}
\end{equation}
For $l_{\rm A} > l_{\rm cr}$ the Lorentz factor $\gamma$
increases with $x$, and the r.h.s. in (\ref{new}) can be omitted.
This gives
\begin{equation}
x^2\frac{\partial \gamma}{\partial x}
+l_{\rm A} x^{2-n}\gamma^2 = C(\theta).
\label{nw}
\end{equation}
Here the integration constant
$C(\theta) \approx l_{\rm A}x_{\rm F}^{2-n}\gamma_{\rm F}^2$
for $l_{\rm A} \gg l_{\rm cr}$ actually does not depend on the
boundary conditions ${\rm d}\gamma/{\rm d}x$ at $x = x_{\rm F}$.
In what follows we do not consider the $\theta$-dependence.

Now analyzing equation (\ref{nw}), one can find that for
$x_{\rm F} \ll x \ll x_{\rm cr}$, where
\begin{equation}
x_{\rm cr} \approx
\left(\frac{l_{\rm A}}{l_{\rm cr}}\right)^{8/(5n-n^2)}x_{\rm F} \ ,
\end{equation}
the particle energy increases as
\begin{equation}
\gamma(x) \approx
\sigma^{1/3}
\left(\frac{x}{x_{\rm F}}\right)^{(n-2)/2}
\end{equation}
for arbitrary $n$. As to the disturbance of magnetic surfaces
$\varepsilon f$, it remains approximately constant:
$\varepsilon f \approx (l_{\rm A}/\sigma)^{2/(5-n)}$.

On the other hand, in the saturation region $x \gg x_{\rm cr}$,
where one can now neglect the term $l_{\rm A}x^{2-n}\gamma^2$ in (\ref{nw}),
we have
\begin{equation}
\gamma(x) =
C\left(\frac{1}{x_{\rm b}}-\frac{1}{x}\right) \ ,
\end{equation}
where $x_{\rm b} \approx x_{\rm cr}$.
One can easily check that the maximum energy
$\gamma_{\rm max} \approx C/x_{\rm cr} \approx
\sigma^{1/3}(x_{\rm cr}/x_{\rm F})^{(n-2)/2}$
corresponds to (\ref{gmax}).
As demonstrated in Fig. 2., the analytical estimate (\ref{gmax})
is in good agreement with the numerical integration of equation (\ref{nw}).

\begin{figure}
\vspace{6cm}
\includegraphics{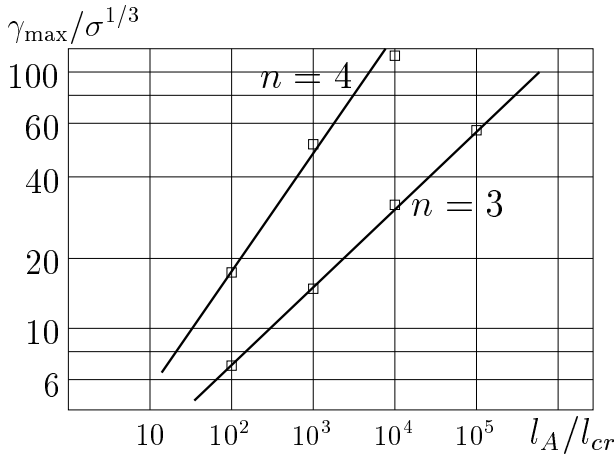}
\caption
{
Saturation energy $\gamma_{\rm max}mc^2$ determined by a
numerical integration of (\ref{nw}) for $n=3$ and $n=4$ (points).
The lines correspond to the analytical estimate (\ref{gmax})
For $l_{\rm A} = l_{\rm max}$, $n\ge 3$ we have
$\gamma_{\rm sup} = \sigma^{7/9}$.}
\end{figure}

\subsubsection{Physical interpretation of particle acceleration}

At first glance, it is quite unexpected that a drag force can result in
an acceleration of particles (so even the word `drag'
itself is not appropriate any more). As was demonstrated,
additional acceleration can be realized for a
highly magnetized outflow when the photon density decreases rapidly
with radius ($n > 2$). This acceleration occurs as a result of the action
of the drag force onto magnetic surfaces.
Thus, acceleration is only obtained when the disturbance of the
magnetic surfaces is taken into account self-consistently.

To understand the nature of the additional particle acceleration
in the supersonic region $r > r_{\rm F}$, it is necessary to return
to the system (\ref{qs1})--(\ref{qs2}).
As one can see, equation (\ref{qs2}) actually plays the role of the GS
equation, describing the force balance in the direction perpendicular to the
magnetic surfaces. It contains no drag term, because
the drag force acts along the magnetic surfaces.
Indeed, as was shown before (Bogovalov 1998;
Okamoto 1999; Beskin \& Okamoto 2000), in the asymptotic region
$r \gg r_{\rm F}$ the transfield GS equation
can be written down as ${\cal F}_{\rm c} = {\cal F}_{\rm em}$, i.e., as
a competition between the centrifugal volume force
\begin{equation}
{\cal F}_{\rm c} = \frac{nmc^2\gamma+S/c}{R_{\rm c}}
\end{equation}
and the electromagnetic volume force
\begin{equation}
{\cal F}_{\rm em} = \rho_{\rm e}E_{\theta} +
\frac{1}{c}j_{\parallel}B_{\varphi}
\approx \frac{1}{8\pi r \sin^2\theta} \, \frac{\partial}{\partial \theta}
\left[(B_{\varphi}^2-E_{\theta}^2)\sin^2\theta\right].
\end{equation}
Here $S \approx (c/4\pi)E_{\theta}B_{\varphi}$ is the Poynting flux.
Now using the expression for the curvature radius $R_{\rm c}$
(Begelman \& Li 1994; Beskin \& Okamoto 2000),
\begin{equation}
\frac{1}{R_{\rm c}} = \frac{\varepsilon}{r\sin\theta}
\frac{\partial}{\partial r}\left(r^2\frac{\partial f}{\partial r}\right),
\label{cur}
\end{equation}
and the condition
\begin{equation}
B_{\varphi}^2-E_{\theta}^2 \approx \frac{1}{\gamma^2}B_{\varphi}^2
\end{equation}
resulting from the relativistic Bernoulli equation
(see e.g. Bogovalov 1998), for a magnetically-dominated flow
$S \gg nmc^3\gamma$ we recover relation (\ref{qs2}).

On the other hand, the Bernoulli equation (\ref{qs1}) describes the change
of the total energy flux (\ref{E}) due to the drag force. As was shown above,
for $l_{\rm A} > l_{\rm cr}$
in the vicinity of the fast magnetosonic surface the leading terms in the
energy equation (\ref{qs1}) are those containing $\varepsilon f$
and $l_{\rm A}$. This means that here the drag force again acts mainly
on the electromagnetic (Poynting) flux $S$. The drag diminishes the
$\theta$-component of the electric field $E_{\theta}$, i.e. it disturbs
the equipotential surfaces $\delta(r, \theta) = $ const. But since in
the one-fluid approximation the magnetic surfaces
$\varepsilon f(r, \theta) =$ const must follow the equipotential surfaces,
the decrease of the electric field $E_{\theta}$ results in a change of
the curvature of magnetic field lines as well. As the condition
${\cal F}_{\rm c} = {\cal F}_{\rm em}$ can now be rewritten in the form
\begin{equation}
\gamma^2 \approx \frac{4\pi j_{\parallel}}{cE_{\theta}}R_{\rm c}
\approx \frac{R_{\rm c}}{r},
\end{equation}
we see that the increase of the curvature radius $R_{\rm c}$
faster than $r$ results in an increase of the particle energy.
In our case such an acceleration is due to the rapid
decrease ($n > 2$) of the isotropic photon density with the distance $r$.

Thus, we see that for high enough photon density
there is an additional acceleration of outgoing plasma
outside the fast magnetosonic surface.
On the other hand, as for $l_{\rm A} > l_{\rm max}$ (\ref{lmax})
the drag force significantly diminishes the total energy flux $E$,
one can conclude that this acceleration may only take place
up to the energy
\begin{equation}
{\cal E}_{\rm sup} \sim \sigma^{1/3}
\left(\frac{l_{\rm max}}{l_{\rm cr}}\right)^{4(n-2)/n(5-n)}m_{\rm e}c^2
\end{equation}
for $2 < n < 3$. On the other hand, ${\cal E}_{\rm sup} \sim
\sigma^{7/9}m_{\rm e}c^2$ for $n\ge 3$.
This energy is always much lower than $\sigma m_{\rm e}c^2$,
corresponding to the total conversion of
electromagnetic energy intoparticle energy.

\section{The Electron-Proton Outflow}

In this section we briefly consider the results of the
analysis of the electron-proton outflow. As the procedure is
quite similar to the electron-positron case, we present
here the principal relations only.

The main difference from the electron-positron case that
occurs due to the large mass ratio is that the drag
force acts on the electron component only, but the
mass flow is determined entirely by protons. As a result,
the energy conservation law has the form
\begin{eqnarray}
\zeta =
\frac{2\varepsilon}{\tan\theta}f
-\frac{(\lambda+\cos\theta/2)\gamma^-
+ (m_{\rm p}/m_{\rm e})(\lambda-\cos\theta/2)\gamma^+}
{2\lambda\sigma\sin\theta}
\nonumber \\
-\frac{l_{\rm A}}{2\lambda\sigma\sin\theta}
\int_{x_0}^x
\left[
\left(
\lambda +\frac{1}{2}\cos\theta
\right)(\gamma^-)^2
+\left(\frac{m_{\rm e}}{m_{\rm p}}\right)^2
\left(
\lambda +\frac{1}{2}\cos\theta
\right)(\gamma^+)^2
\right]
u(x^{\prime})\gamma^2(x^{\prime}){\rm d}x^{\prime}
\nonumber \\
\approx \frac{2\varepsilon}{\tan\theta}f
-\left(\frac{m_{\rm p}}{m_{\rm e}}\right)
\frac{\gamma}{2\sigma\sin\theta}
-\frac{1}{2\sigma\sin\theta}l_{\rm A}\int_{x_0}^xu(x^{\prime})
\gamma^2(x^{\prime})
{\rm d}x^{\prime},
\label{enp}
\end{eqnarray}
where now $\gamma = \gamma^+$.
The conservation of angular momentum
(\ref{mom1})--(\ref{mom2}) looks like
\begin{eqnarray}
\gamma^+(1-x\sin\theta\xi_{\varphi}^+) & = &
\gamma_{\rm in}^+ -l_{\rm A}\left(\frac{m_{\rm e}}{m_{\rm p}}\right)^3
\int_{x_0}^xu(x^{\prime})
(1-x^{\prime}\sin\theta\xi_{\varphi}^+)(\gamma^+)^2{\rm d}x^{\prime} -
4\lambda\sigma\frac{m_{\rm e}}{m_{\rm p}}(\delta - \varepsilon f),
\label{mom1a}  \\ \gamma^-(1-x\sin\theta\xi_{\varphi}^-) & = &
\gamma_{\rm in}^- -l_{\rm A}\int_{x_0}^xu(x^{\prime})
(1-x^{\prime}\sin\theta\xi_{\varphi}^-)(\gamma^-)^2{\rm d}x^{\prime} +
4\lambda\sigma(\delta - \varepsilon f),
\label{mom2a}
\end{eqnarray}
resulting in $\xi_{\varphi} = 1/x\sin\theta$.
Since for $r \ll r_{\rm F}$ we again have
$\xi_r = \xi_{\varphi}/(x\sin\theta)$,
we return to the universal dependence
\begin{equation}
\gamma \approx x \sin\theta.
\label{x}
\end{equation}

Equation (\ref{m_gamma}) determining the energy on the fast
magnetosonic surface now has the form
\begin{eqnarray}
\gamma^3 - 2\sigma\frac{m_{\rm e}}{m_{\rm p}}
\left[2\varepsilon f\cos\theta -
\varepsilon\sin\theta\frac{\partial f}{\partial\theta}
-\frac{1}{2}\frac{l_{\rm A}}{\sigma}\int_{x_0}^x
u(x^{\prime})\gamma^2(x^{\prime})
{\rm d}x^{\prime} +\frac{1}{2x^2}\right]\gamma^2
+ \frac{m_{\rm e}}{m_{\rm p}}\sigma\sin^2\theta=0.
\label{m_gammap}
\end{eqnarray}
Hence, for arbitrary $l_{\rm A}$
the Lorentz factor of particles can be presented in the form
\begin{equation}
\gamma_{\rm F} =  \left(\frac{2m_{\rm e}}{m_{\rm p}}\right)^{1/3}
\sigma^{1/3}\sin^{2/3}\theta \ .
\end{equation}
Now, for $l_{\rm A} < l_{\rm cr}^{(\rm p)}$
the position of the fast magnetosonic surface and the disturbance
of magnetic surfaces on the fast magnetosonic surface are
\begin{eqnarray}
x_{\rm F} & \approx & \left(\frac{2m_{\rm e}}{m_{\rm p}}\right)^{1/3}
\sigma^{1/3} \sin^{-1/3}\theta \ , \\
(\varepsilon f)_{\rm F} & \approx & \left(\frac{2m_{\rm e}}
{m_{\rm p}}\right)^{-2/3} \sigma^{-2/3}.
\end{eqnarray}
Here for $n < 3$ we have
\begin{equation}
l_{\rm cr}^{(\rm p)}
= \left(\frac{m_{\rm p}}{m_{\rm e}}\right)^{(5-n)/3}\sigma^{(n-2)/3},
\end{equation}
and for $n \ge 3$ we have
$l_{\rm cr}^{(\rm p)} = (m_{\rm p}/m_{\rm e})^{2/3}\sigma^{1/3}$.
On the other hand, for $l_{\rm A} > l_{\rm cr}^{(\rm p)}$ we return
to the relations $x_{\rm F} = (\sigma/l_{\rm A})^{1/2}$,
$(\varepsilon f)_{\rm F} = l_{\rm A}/\sigma$ for $n \ge 3$,
but for $n < 3$ now
\begin{eqnarray}
x_{\rm F} & \approx &
\left(\frac{\sigma}{l_{\rm A}}\right)^{1/(5-n)},
\\
(\varepsilon f)_{\rm F} & \approx &
\left(\frac{l_{\rm A}}{\sigma}\right)^{2/(5-n)}.
\end{eqnarray}
As we see, for the disturbance of magnetic surfaces
and the position of the fast magnetosonic surface
remain the same as for the electron-positron outflow.
In particular, the terminating compactness parameter
$l_{\rm max} = \sigma$ is again determined by electrons.

\begin{figure}
\vspace{5.5cm}
\includegraphics{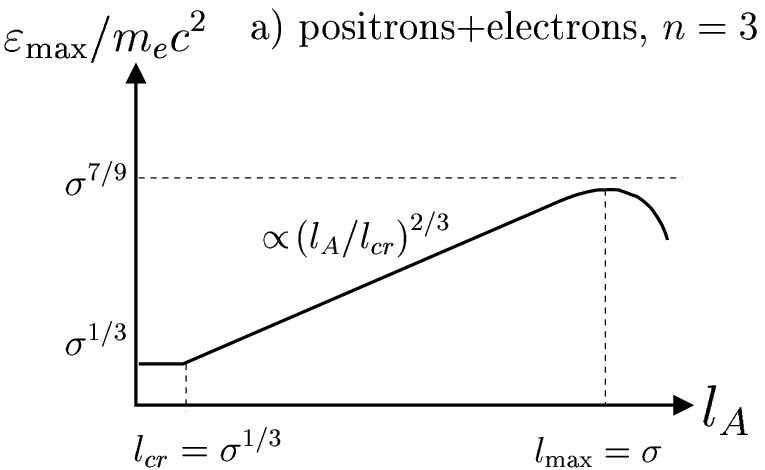}
\includegraphics{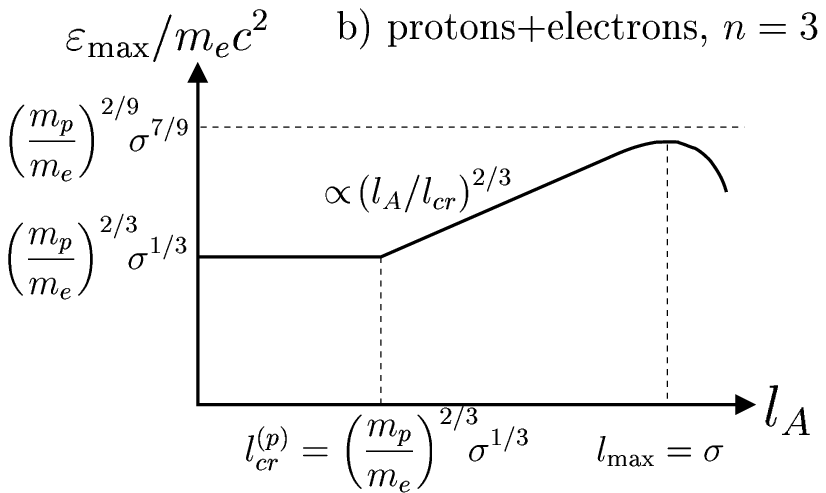}
\caption
{Dependence of the maximum particle energy
${\cal E}_{\rm max}$
on the compactness parameter $l_{\rm A}$
both for electron-proton ($a$) and
electron-positron ($b$) outflows for $n=3$.
Here $m_{\rm p}$ is a proton mass.
The decrease of the particle energy for a very large
compactness parameter $l_{\rm A}$ is due to the
decrease of the total energy flux $E_{\rm B}$.
}
\end{figure}

Thus, one can conclude that inside the fast magnetosonic surface
$r < r_{\rm F}$ the Lorentz factors of all particles are given
by the universal relation (\ref{x}). Hence, our results demonstrate that
the drag force does not affect the structure of the flow inside the fast
magnetosonic surface for both the electron-positron and the electron-proton
cases. But the position of the fast magnetosonic surface and the critical
value of the photon density depend on the mass of the outgoing particles,
and for the electron-proton case they are determined by protons.

At large distances $r > r_{\rm F}$ for $l_{\rm A} < l_{\rm cr}^{(\rm p)}$
the particle energy actually remains the same as on the fast
magnetosonic surface ($\gamma(r > r_{\rm F}) \approx \gamma_{\rm F}$),
while for $l_{\rm A} > l_{\rm cr}^{(\rm p)}$ the Lorentz factor
increases up to the maximum value
\begin{equation}
\gamma_{\rm max} \approx \gamma_{\rm F}
\left(\frac{l_{\rm A}}{l_{\rm cr}^{(\rm p)}}\right)^{4(n-2)/(5n-n^2)}
\end{equation}
for $n < 3$.
This takes place up to the distance $r \approx r_{\rm cr}$, where
\begin{equation}
r_{\rm cr} = r_{\rm F}
\left(\frac{l_{\rm A}}{l_{\rm cr}^{(\rm p)}}\right)^{8/(5n-n^2)} \ .
\end{equation}
Clearly, this additional acceleration can be realized
only for $\sigma > m_{\rm p}/m_{\rm e}$.
Then, we have for $n = 3$
\begin{equation}
{\cal E}_{\rm sup} \sim \left(\frac{m_{\rm p}}{m_{\rm e}}\right)^{2/9}
\sigma^{7/9}m_{\rm e}c^2\ .
\end{equation}

\section{Discussion and Astrophysical Applications}

We have demonstrated how for a simple geometry
it is possible to determine a small radiation-drag-force correction to
the one-fluid ideal MHD outflow.
The disturbance of magnetic surfaces was self-consistently taken into
consideration. As a result, it is
possible to characterize the general influence of the drag action on the
magnetic field structure for an ideal magnetically-dominated
quasi-monopole cold outflow
and to determine under what circumstances radiation drag is important.

As demonstrated above, the characteristics of the flow are
determined by two main parameters, namely
the compactness parameter $l_{\rm A}$ (\ref{A}) (which is proportional
to the photon density)
and the magnetization parameter $\sigma$ (\ref{sigma}).
If the photon density is low, so that
the compactness parameter is small $l_{\rm A} \ll l_{\rm cr}(\sigma)$,
the action of the drag force is negligible,
while for a high photon density $l_{\rm A} \gg l_{\rm cr}(\sigma)$,
particles are additionally accelerated
outside the fast magnetosonic surface.

In particular, for $l_{\rm cr} \ll l_{\rm A} \ll l_{\rm max}$
the increase of the drag force results in an increase
of the outgoing plasma energy
${\cal E}_{\rm max} \approx \gamma_{\rm max}m_{\rm e, p}c^2$,
but the disturbance of magnetic surfaces is small ($\varepsilon f \ll 1$).
For $l_{\rm A} \sim l_{\rm max}$
an increase of the photon density results in the increase
of collimation up to values $\varepsilon f \sim 1$,
but the particle energy remains
near the saturation value ${\cal E}_{\rm sup}$.
Finally, for a very high photon density $l_{\rm A} \gg l_{\rm max}$
an effective collimation of magnetic surfaces becomes possible,
but in this case the drag force substantially diminishes
the flux of electromagnetic energy inside the fast magnetosonic
surface. As a result, for $l_{\rm A} \gg l_{\rm max}$ almost all the
energy of the electromagnetic field is lost via the
inverse Compton interaction  of particles with external photons.
For this reason, the very existence of a magnetically-dominated
flow becomes impossible.
The dependence of the maximum particle energy ${\cal E}_{\rm max}
= \gamma_{\rm max}m_{\rm e, p}c^2$
on the compactness parameter $l_{\rm A}$ is shown in Fig. 3.

We now consider several astrophysical applications.

\subsection{Active Galactic Nuclei}

For AGNs (the central engine is assumed to be a rotating black hole with
mass $M \sim 10^9M_{\odot}$, $R \sim 10^{14}$ cm,
the total luminosity $L \sim 10^{45}$ erg s$^{-1}$, $B_0 \sim 10^4$ G)
the compactness parameter $l_{\rm A}$ (\ref{A}) can be evaluated as
\begin{equation}
l_{\rm A} \approx 30M_9^{-1}\left(\frac{\Omega R}{c}\right)L_{45}.
\end{equation}
In the Michel magnetization parameter $\sigma$ (\ref{sigma})
\begin{equation}
\sigma \approx 10^{14}\lambda^{-1}M_9B_4\left(\frac{\Omega R}{c}\right),
\end{equation}
the main uncertainty comes from the multiplication parameter $\lambda$,
i.e., in the particle number density $n$. Indeed, for an electron-positron
outflow this value depends on the efficiency of pair creation
in the magnetosphere of a black hole, which is still undetermined.
In particular, this process depends on the density and
energies of the photons in the immediate vicinity of the black hole.
As a result, if the hard-photon density is not high, then the multiplication
parameter is small ($\lambda \sim 10 - 100$; Beskin, Istomin \&
Pariev 1992; Hirotani \& Okamoto 1998). In this case for
$(\Omega R/c) \sim 0.1$--$0.01$ we have
$\sigma \sim 10^9 - 10^{12}$, so that $l_{\rm cr} \sim 10^3$--$10^4$.
On the other hand, if the density of photons
with energies ${\cal E}_{\gamma} > 1$MeV is high enough,
direct particle creation $\gamma + \gamma \rightarrow e^+ +e^-$
results in an increase of the particle density (Svensson 1984). This gives
$\sigma \sim 10 - 10^3$, and hence
$l_{\rm cr} \sim 10$ for an electron-positron outflow.

From a theoretical point of view, the most interesting result here is the
possibility of an additional acceleration of particles
outside the fast magnetosonic surface. Indeed, for a high enough
photon density ($l_{\rm A} \sim 10 - 100$, i.e., for
$L \sim 10^{46} - 10^{48}$ erg s$^{-1}$) and a small magnetization
parameter $\sigma \sim 10 - 100$,
the compactness parameter $l_{\rm A}$ can exceed the critical value
$l_{\rm cr}$ for an electron-positron outflow. In this case, according to Fig. $1b$, our analysis suggests that the kinetic luminosity of the relativistic jet should be
proportional to $l_{\rm A}^{2/3}\propto L_{\rm tot}$, where
$L_{\rm tot}$ in the total luminosity of the central engine. Kinetic luminosity is not easily determined from observations. However, observational evidence suggests that the radio luminosity of the jets is positively correlated with the luminosity of the central engine and the scatter of this correlation decreases towards larger luminosities (Baum, Zirbel \& O'Dea 1995).

For an electron-proton outflow the magnetization parameter (\ref{sigma})
can be rewritten in the form (Camenzind 1990)
\begin{equation}
\sigma = \frac{m_{\rm p}}{m_{\rm e}}\left(\frac{\Omega R}{c}\right)^2
\frac{B_0^2R^2}{c{\dot M}}
\approx  3 \times 10^4 \left(\frac{\Omega R}{c}\right)^2
B_4^2M_9^2\left(\frac{{\dot M}}{0.1\,M_{\odot}/{\rm yr}}\right)^{-1}.
\end{equation}
Here ${\dot M} = 4\pi nm_{\rm p}R^2c$ is the mass ejection rate.
Hence, for a high ejection rate (${\dot M} > 0.1\,M_{\odot}$~yr $^{-1}$) the
magnetization parameter $\sigma < m_{\rm p}/m_{\rm e}$. In this
case there is no acceleration of plasma.
On the other hand, for low ejection rate ${\dot M} < 0.1\,M_{\odot}/{\rm yr}$
the magnetization parameter becomes too large for the drag force
to be efficient.

Thus, the drag force can substantially disturb
the MHD parameters of a Poynting-dominated outflow
only for a very high luminosity of the central engine
($L_{\rm tot} \gg 10^{45}$~erg~s $^{-1}$) and only for
an electron-positron outflow.
In all other cases the action of the drag force remains negligible.
In particular, the additional acceleration of particles outside
the fast magnetosonic surface is not efficient.

\subsection{Cosmological Gamma-Ray Bursts}

For cosmological gamma-ray bursts (the central engine is represented by
the merger of very rapidly orbiting neutron stars or black holes
with $M \sim M_{\odot}$, $R \sim 10^{6}$ cm,
total luminosity
$L \sim 10^{52}$ erg s$^{-1}$, $B_0 \sim 10^{15}$ G;
see, e.g., Lee et al 2000 for details)
the compactness parameter $l_{\rm A}$ is extremely large:
\begin{equation}
l_{\rm A} \sim 10^{17}\left(\frac{\Omega R}{c}\right)L_{52}.
\end{equation}
On the other hand, even for a superstrong magnetic field of
$B_0 \sim 10^{15}$ G (which is necessary to explain the
total energy release) the magnetization parameter $\sigma$
is small ($\sigma < 1 - 10$), because within this model the
magnetic field itself is secondary and its energy density
cannot exceed the plasma energy density.
Thus, one can conclude that for these characteristics of
cosmological gamma-ray bursts the density of photons is very high
so that $l_{\rm A} \gg l_{\rm max}$ and the drag force can make
it difficult to form a Poynting-dominated outflow. A self-consistent
analysis should include into consideration other physical processes
such as high optical thickness resulting in the diminishing of the
photon density, radiation and particle pressure, {\it etc.}
Nevertheless, in our opinion, our conclusion
may substantially restrict some recent models of
cosmological gamma-ray bursts.

\subsection{Radio Pulsars}
For radio pulsars the central engine is a rotating neutron star with
$M \sim M_{\odot}$, $R \sim 10^{6}$ cm,
total luminosity of the surface $L_{\rm X} \sim 10^{33}$--$10^{37}$
erg s$^{-1}$,
and $B_0 \sim 10^{12}$ G. In this case the magnetization parameter
$\sigma \sim 10^4$--$10^6$, corresponding to relativistic electron-positron
plasma, is known with rather high accuracy (see, e.g., Bogovalov 1997).
This gives $l_{\rm cr} \sim 10^2$--$10^3$, and the compactness parameter
\begin{equation}
l_{\rm A} \sim \left(\frac{\Omega R}{c}\right)L_{35}
\end{equation}
remains small ($< 1$) even for the most energetic
($L_{\rm X} \sim 10^{37}$ erg s $^{-1}$ ) fast ($\Omega R/c \sim 10^{-2}$)
pulsars like Crab and Vela. Thus, one can conclude that
the drag force does not substantially disturb the
magnetically-dominated outflow from radio pulsars.

Thus, the drag force does not affect the wind characteristics (particle energy, magnetic field structure, etc.) of pulsars. However, interaction of outflowing relativistic particles with thermal photons can be important in other ways. In the wind region ($r \gg R_{\rm L}$) even a weak interaction with photons can result in a detectable flux of inverse Compton gamma-ray photons (Bogovalov \& Aharonian 2000). On the other hand, near the surface of the star ($r \ll R_{\rm L}$), inverse Compton photons are important in the pair creation process (Kardashev, Mitrofanov \& Novikov 1984; Zhang \& Harding 2000).

\section*{Acknowledgments}

The authors are grateful to K.A.Postnov for fruitful discussions.
We thank the referee for a detailed report and useful comments.
VSB thanks Observatory Paris-Meudon for hospitality.
This work was partially supported by Russian Foundation
for Fundamental Research (Grant 02-02-16762).

{}

\end{document}